\documentclass[fleqn,usenatbib,usedcolumn]{mnras}

\let\la=\lesssim % for less than similar from newtxmath, not \la from  mnras.cls
\usepackage{amsmath}	% Advanced maths commands
\usepackage{amssymb}	% Extra maths symbols

\usepackage{txfonts}

\usepackage[T1]{fontenc}
\usepackage{graphicx}	% Including figure files

\newcommand{\Msun}{\>{\rm M_{\odot}}}

\title[The origin of type-I profiles in cluster lenticulars]{The origin of type-I profiles in cluster lenticulars: An interplay between ram pressure stripping and tidally-induced spiral migration}

\author[Clarke et.\ al.]{
Adam J. Clarke,$^{1}$\thanks{E-mail: aclarke13@uclan.ac.uk}
Victor P. Debattista,$^{1}$
Rok Ro\v{s}kar$^{2}$
and Tom Quinn$^{3}$\\
$^1$Jeremiah Horrocks Institute, University of Central Lancashire, Preston, PR1
2HE, UK \\
$^2$ Research Informatics, Scientific IT Services, ETH Z\"urich,
Weinbergstrasse 11, 8092, Z\"urich, Switzerland \\
$^3$ Astronomy Department, University of Washington, Box 351580, Seattle, WA
98195, USA
}

\date{Accepted 2016 October 11.\@ Received 2016 October 11;\@ in original form 2016 July 28}

\pubyear{2016}

\begin{document}
\label{firstpage}
\pagerange{\pageref{firstpage}--\pageref{lastpage}}
\maketitle

\begin{abstract}
Using $N$-body+SPH simulations of galaxies falling into a cluster, we
study the evolution of their radial density profiles.  When evolved in
isolation, galaxies develop a type~II (down-bending)
profile.  In the cluster, the evolution of the profile depends on the
minimum cluster-centric radius the galaxy reaches, which controls the
degree of ram pressure stripping.  If the galaxy falls to $\sim 50\%$
of the virial radius, then the profile remains type~II, but if the
galaxy reaches down to $\sim 20\%$ of the virial radius the break
weakens and the profile becomes more type~I like. The velocity
dispersions are only slightly increased in the cluster simulations
compared with the isolated galaxy; random motion therefore cannot be
responsible for redistributing material sufficiently to cause the
change in the profile type.  Instead we find that the joint action of
radial migration driven by tidally-induced spirals and the outside-in
quenching of star formation due to ram pressure stripping alters the
density profile.  As a result, this model predicts a flattening of the
age profiles amongst cluster lenticulars with type~I profiles, which
can be observationally tested.
\end{abstract}


%%%%%%%%%%%%%%%%%%%%%%%%%%%%%%%%%%%%%%%%%%%%%%%%%%

%%%%%%%%%%%%%%%%% BODY OF PAPER %%%%%%%%%%%%%%%%%%

\section{Introduction}

Lenticular (S0) galaxies share the discy morphology of spirals, but
are red and dead, with a lack of young stars and a smooth appearance
which results from an absence of a substantial amount of cold gas
\citep[e.g.][]{Chamaraux1986}. They tend to be found more
frequently in clusters, while the field environment favours spiral
galaxies \citep{dressler1980, cappellari2011}. The fraction of
lenticulars decreases with redshift, while that of spirals increases
\citep{dressler1997, couch1998, postman2005}, implying that lenticular
galaxies form via the transformation of spirals. Various mechanisms have
been proposed to explain the quenching of their star formation (SF).
Ram pressure stripping (RPS) can remove the majority of cool
gas \citep{gunn1972, quilis2000} or the hot gas
corona, preventing future gas cooling onto the disc and ending a
galaxy's ability to form stars, often referred to as
strangulation \citep{larson1980, bekki2002, steinhauser2016}.
Harassment \citep{moore1996, moore1999} and minor-merger triggered
starbursts leading to gas depletion \citep{mihos1994, bekki1998}
can also quench SF. It is not yet clear which of these
mechanisms dominates.

It has long been known that the bright parts of disc galaxies show an
exponentially declining surface density \citep{freeman1970}.  However
most profiles have a change in scale length at a ``break-radius''.
\citet{freeman1970} defines two disc profile types.  The first, type~I
profiles, are purely exponential out to the last measured point
\citep[for example in NGC~300 this extends to 10~scale
  lengths][]{bland-hawthorn2005, erwin2008, vlajic2011}.  Type~II
profiles instead show a decrease in scale-length past a
break-radius. These breaks have been shown to originate from a drop in
the star formation rate (SFR) \citep{schaye2004, roskar2008a,
  radburn-smith2012}.  The outer disc is then populated by stars
migrating outward due to transient spiral structure, via the
corotation resonance trapping mechanism proposed by
\citet{sellwood2002}.  This leads to a type~II profile with a break
radius that moves outwards as gas cools onto the disc
\citep{roskar2008a}.  Alternatively type~II profiles have also been
explained without a SF threshold, via angular momentum
exchanges mediated by bars and spirals \citep{debattista2006,
  foyle2008, minchev2010} although this leads to hot outer
discs.  Galaxies with type~II profiles exhibit an upturn in their
colour profile close to the break radius \citep{bakos2008,
  azzollini2008}. A third type of profile, termed type~III, shows an
increase in scale-length past the break-radius \citep{erwin2005}.
Type~III profiles are generally thought to result from heating of the
inner disc through various mechanisms \citep[e.g.][]{younger2007,
  minchev2012, borlaff2014, herpich2015}.

\citet{erwin2012} studied the distribution of light profiles amongst
lenticular galaxies. They found that the fraction of type~I and II
profiles depends on the environment.  Amongst the field lenticulars
type~I and type~II profiles were equally common.  However they found
virtually no type~II profiles amongst cluster lenticulars and double
the frequency of type~I profiles.  \citet{roediger2012} also found no
type~II profiles amongst Virgo lenticulars. \citet{gutierrez2011}
found that 33\% of their sample of lenticulars, but only 10\% of
spirals, showed a type~I profile, and that the frequency of type~II
profiles changes from 80\% to 25\% for the spiral and lenticular
samples, respectively.~\cite{pranger+16} constructed two samples of
galaxies with mass $1-4\times 10^{10}$M$_{\odot}$, one in the field and
one in the cluster environment.  Comparing the profiles, they found
that type~I profiles are three times more frequent in clusters than in
the field.  Because they considered the outer discs, at surface
brightness $24 < \mu < 26.5$ mag.\ arcsec$^{-2}$, beyond where most
disc breaks are located \citep{pohlen2006}, \citet{maltby2015} found
remarkably similar frequencies of type~I/II/III profiles amongst
lenticulars in the cluster and field environments.  They also found no
difference in disc scale-lengths between lenticular and spiral
galaxies.  These studies imply that environmental processes driving
the evolution of lenticulars may also be responsible for the
light-profile properties, which may provide insight into which cluster
lenticular formation mechanism is most important.~\cite{herpich2015}
demonstrated that pure exponential discs occur in simulations
when the halo has a narrow range of angular momentum, potentially explaining
how isolated galaxies can exhibit single-exponential profiles.

In this Letter, we use $N$-body+SPH simulations of spiral galaxies
falling into a cluster where they experience RPS,
substantially quenching their SF.  We consider the
evolution of the density profiles compared with the same galaxy
evolved in isolation and demonstrate that the mechanism driving the
differences is radial migration enhanced by tidally-induced spirals,
with RPS playing a vital role by quenching star
formation outside-in.

\section{Simulations}

Our simulations consist of single galaxies falling into a cluster. The
initial conditions for the galaxy are the same as those used in the
well studied system of \citet{roskar2008a, roskar2008b} and
\citet{loebman2011}, with a spherical NFW dark matter halo
\citep{navarro1995} and an embedded spherical corona of gas with a
temperature profile such that hydrostatic equilibrium is established.
We impart an angular momentum, $j \propto R$, to the gas, with a spin
parameter of $\lambda = 0.065$ \citep{Bullock2001}.  The dark matter
consists of two shells, the inner containing $9 \times 10^5$ particles
of mass $10^6$~M$_{\odot}$ extending to 200~kpc and the outer
containing $1 \times 10^5$ particles of mass $3.5 \times
10^6$~M$_{\odot}$.  There are $10^6$ gas particles, each with mass
$1.4 \times 10^5~$M$_{\odot}$.  The total mass within the virial
radius (R$_{200}$ = 200~kpc) is $10^{12}$~M$_{\odot}$. We use a
softening of 50~pc for the gas and stars, and 100~pc for the dark
matter.

We consider a cluster similar in size and mass to the Fornax cluster.
The virial radius is set to $0.7$~Mpc and the virial mass enclosed is
$6 \times 10^{13}$~M$_{\odot}$ \citep{ikebe1992, drinkwater2001,
  nasonova2011}.  We model the cluster with $9 \times 10^6$ dark
matter particles of mass $4 \times 10^6$~M$_{\odot}$ in the inner
shell, extending to 700~kpc, and $10^6$ dark matter particles of mass
$2 \times 10^7$~M$_{\odot}$ in the outer shell. There are $2 \times
10^7$ gas particles in the cluster, each of mass $2.3 \times
10^5$~M$_{\odot}$ producing a mass resolution comparable to that in
the galaxy itself.  The softening lengths are set to match the
infalling galaxy.  We prevent the cluster gas from cooling, to mimic
the episodic AGN feedback which prevents substantial SF in
the centres of massive clusters \citep{binney2004}.

We initially place the galaxies at three times the cluster virial
radius, to allow the galaxy to form a disc before RPS commences.  Here
we present just two simulations, which we
refer to as `cluster350' and `cluster150', targeting periapsis at
350~kpc and 150~kpc, respectively. With the cluster centred on the
origin, we place the galaxies along the $z$-axis from the cluster
centre, with the internal angular momentum vector also along the
$z$-axis and the orbital angular momentum vector perpendicular to it,
i.e.\ along the $x$-axis. We give the cluster150 model
velocities $v_y = 273$~km/s and $v_z = 157$~km/s, whilst
the cluster350 model has $v_y = 157$~km/s and $v_z = 273$~km/s.
We use the isolated simulation of
\citet{loebman2011} as a control model to disentangle the effect of
environment and refer to this simulation as the ``isolated'' model.

We evolve the models for 10 Gyr with the $N$-body + smooth particle
hydrodynamics (SPH) code GASOLINE \citep{wadsley2004}.  We adopt star
formation criteria where the gas density and temperature have to be
greater than 0.1 cm$^{-3}$ and less than $15,000$~K,
respectively. Feedback from both Type~II and Type~Ia supernovae is
introduced via energy injected into the interstellar medium in the
form of a sub-grid modelled blast-wave as described in
\citet{stinson2006}.  Stars form with $1/3$ of the gas particle mass,
corresponding to $4.6 \times 10^4$ M$_{\odot}$, and each gas particle
can form multiple stellar particles. The minimum gas mass is set at
$1/5$ of its original mass. Once a gas particle drops below this mass
it is removed, and its mass is distributed to the surrounding gas
particles.  We adopt a base time step of $\Delta t = 0.01$ Gyr,
refining our timesteps using $\delta t = \Delta t / 2^n <
\eta{(\epsilon / a_g)}^{1/2}$, where $\epsilon$ is the softening
length and $a_g$ is the particle acceleration at the current
position. We set the refinement parameter $\eta = 0.175$. The tree
code opening angle $\theta = 0.7$. The timestep for gas particles
also satisfies $\delta t_{\textrm{gas}} = \eta_{\textrm{courant}}h /
[(1 + \alpha)c + \beta \mu_{\max}]$, where $h$ is the
SPH smoothing length, $\eta_{\textrm{courant}} = 0.4$,
 $\alpha = 1$ is the shear coefficient, $\beta = 2$ is the viscosity
 coefficient and
$\mu_{\max}$ is described in~\cite{wadsley2004}.
The SPH kernel is defined using the
nearest 32 neighbours. These parameters have been shown to lead to
realistic late-type galaxies \citep{roskar2012, roskar2013}.

\section{Results}

\begin{figure}
  \centering
  \includegraphics[width=0.99\columnwidth]{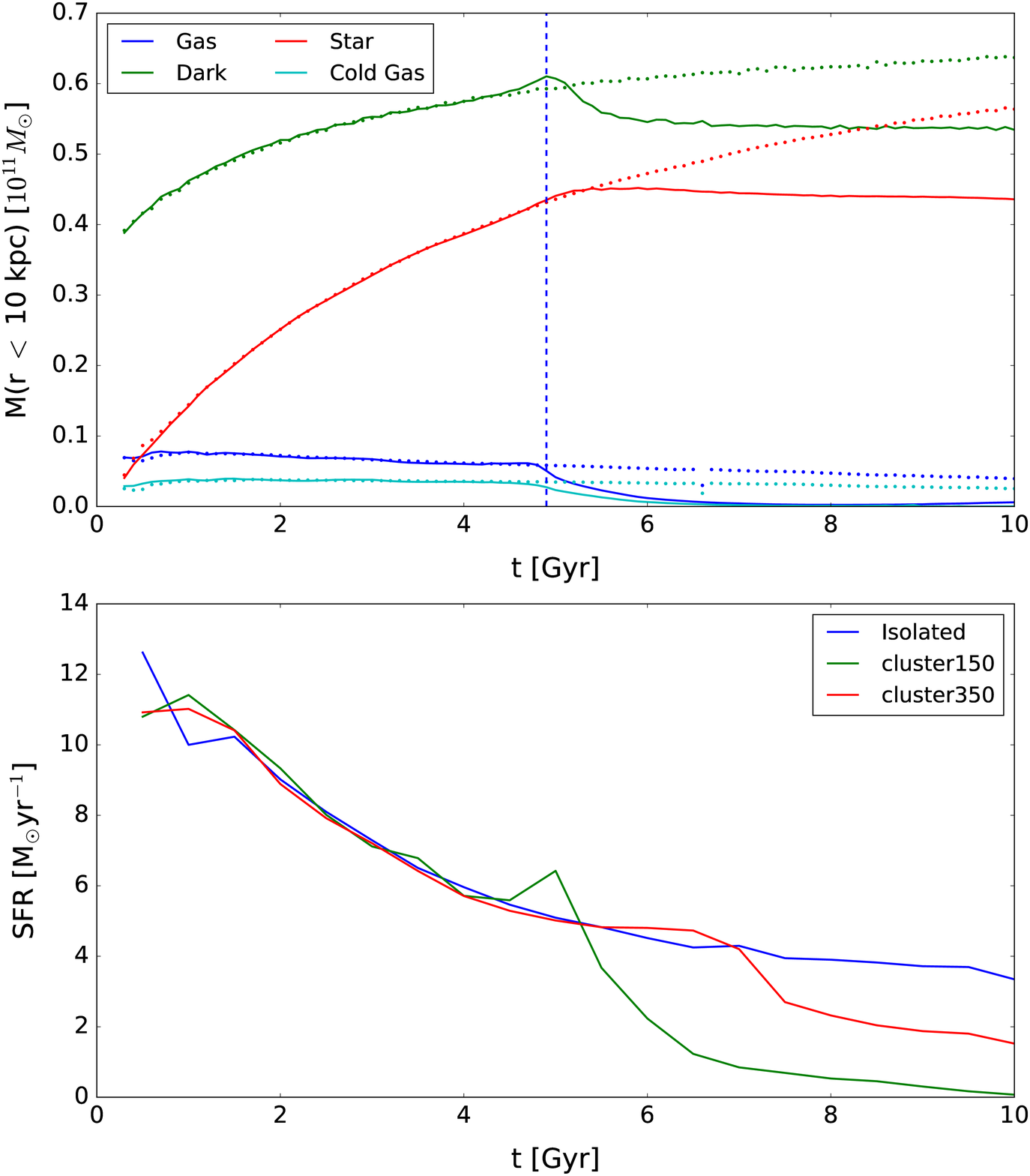}
  \caption{ {\it Top\/}: Evolution of the mass enclosed within the
    inner 10~kpc in the cluster150 galaxy (solid) compared with the
    isolated galaxy (dotted). The blue line shows all gas particles
    whilst the cyan line shows only cool (T \textless{} 15,000 K)
    gas. Green and red show the dark matter and stellar content
    respectively. The vertical lines correspond to the time at which
    the cluster150 galaxy is at periapsis. {\it Bottom\/}: Star
    formation histories for our models, as detailed at top right.}
\label{fig:massenc}
\end{figure}

In the top panel of Figure~\ref{fig:massenc} we show the evolution of
the mass within the inner 10~kpc of the cluster150 model compared with
the isolated galaxy.  Initially, the cluster150 and isolated models
evolve in parallel.  At $\sim 4$ Gyr the galaxy reaches a location in
the cluster dense enough for RPS of the cool gas to
become strong, which occurs around the time of periapsis.  The bottom
panel shows the SF histories for all three models.  In the
cluster galaxies, prior to significant mass loss, there is a burst of
SF, which we attribute to compression of the cool gas by
the cluster medium.  This is strongest in the cluster150 model, but a
milder one is also apparent in the cluster350 model. SF in
the latter model is quenched much more gently, but by 10~Gyr it is
forming stars at roughly half the rate ($\sim2 \Msun yr^{-1}$) of the
isolated model. The cluster150 model terminates SF almost
entirely, except in the very inner ($\lesssim 1$~kpc) regions.  A
small ($\lesssim 10 \%$) fraction of dark matter is also lost from the
inner 10~kpc.

\begin{figure}
  \centering
  \includegraphics[width=0.99\linewidth]{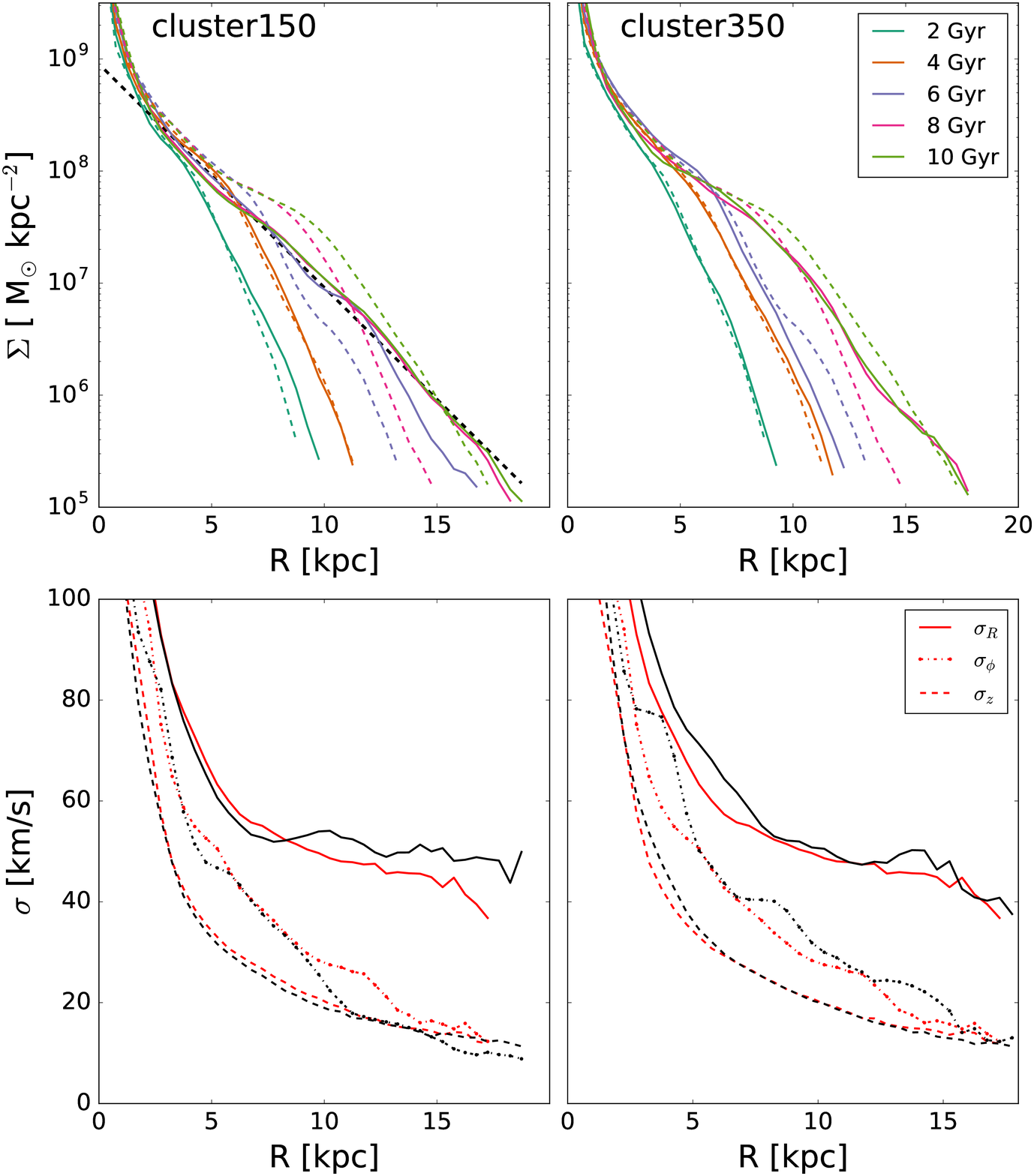}
  \caption{{\it Top\/}: Evolution of the surface density profiles for
    the cluster150 (left) and cluster350 (right) models (solid lines)
    compared with the isolated model (dashed lines) as detailed on the
    right. The black dashed line shows a single-exponential fit to the
    disc component of the 10~Gyr cluster150 model.
    {\it Bottom\/}: Final velocity dispersion profiles for the
    cluster galaxies (black lines) compared with the isolated
    simulation (red lines).}
\label{fig:dispersions_profiles}
\end{figure}

In the top panels of Figure~\ref{fig:dispersions_profiles} we compare
the evolution of the surface density profiles of the isolated and the
cluster galaxies. The isolated galaxy develops a type~II profile due
to the SF drop and outwardly migrating stars moving past
the break radius \citep{roskar2008a}. At early times the cluster150
galaxy (left panel) exhibits a type~II profile, with the break moving
outwards, initially evolving identically to the isolated
galaxy. Once the galaxy is approaching
periapsis and RPS is strongest, it begins
to transition from a type~II to a type~I profile.  It is noteworthy
that in the final profiles the main difference between the isolated
and cluster150 models is not at large radii, but at intermediate
($\sim 7-10$~kpc) radii.  At 8~Gyr, the outer disc of the cluster350
model is more massive than that of the isolated model.  By 10~Gyr the
break in the profile is weaker, but it retains a type~II profile.

\begin{figure}
  \centering
  \includegraphics[width=0.99\columnwidth]{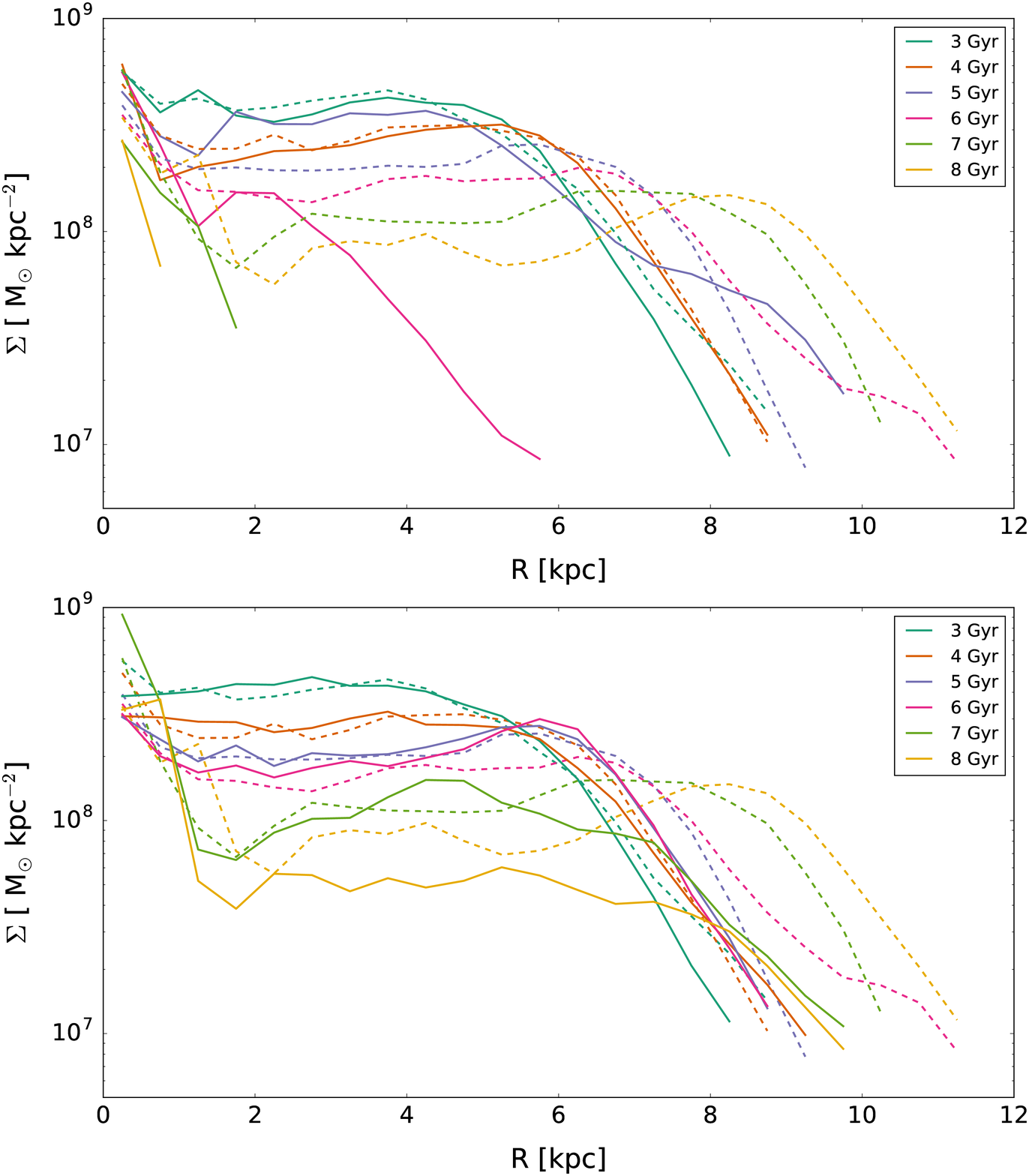}
  \caption{Surface density of young stars (age $<0.5$~Gyr) as a
    function of formation radius for the cluster150 model (top panel,
    solid lines), cluster350 model (bottom panel, solid lines) and the
    isolated model (dashed lines in both panels) at different times
    as detailed by the insets.}
\label{fig:sf_profile}
\end{figure}

Figure~\ref{fig:sf_profile} shows the surface density profiles of
young stars (age $<0.5$~Gyr) as a function of radius. In the isolated
model SF grows inside-out as more gas cools onto the
disc. In comparison, the cluster150 model grows inside-out only up to
the onset of RPS, at which point the SF
quenches from the outside in.  Significant SF never
extends beyond $\sim 6$ kpc; the disc therefore never has an
opportunity to form a substantial mass beyond this point.  This
explains why the difference between the profiles of the cluster150 and
isolated models is largest at these intermediate radii, i.e.\ between
the maximum extent of SF in the cluster150 model and the
final one in the isolated model.  At even larger radii, the relative
difference between the two models is smaller.  In the cluster350
model, starting from about 6~Gyr, the SF break stays
roughly constant while the SFR is slowly reduced.  No
significant amount of SF ever occurs outside 6~kpc in this
model.  Therefore the presence of stars at $R\sim 15$ kpc in both
cluster models requires an explanation.

The bottom panels of Figure~\ref{fig:dispersions_profiles} compare the
mass-weighted velocity dispersions of the cluster galaxies (black) and
the isolated galaxy (red).  The cluster galaxies are not substantially
hotter, in any direction.  For this $\sigma_R$, we expect epicyclic
excursions to be no larger than $\simeq 3$~kpc, since
$\Delta R \simeq \sqrt 2 \sigma_R / \kappa$ and
taking $\kappa = 25$~km~s$^{-1}$~kpc$^{-1}$.
Thus the presence of stars at
large radii in both the cluster150 and the cluster350 models is not
due to heating.

\begin{figure}
  \centering
  \includegraphics[width=0.99\linewidth]{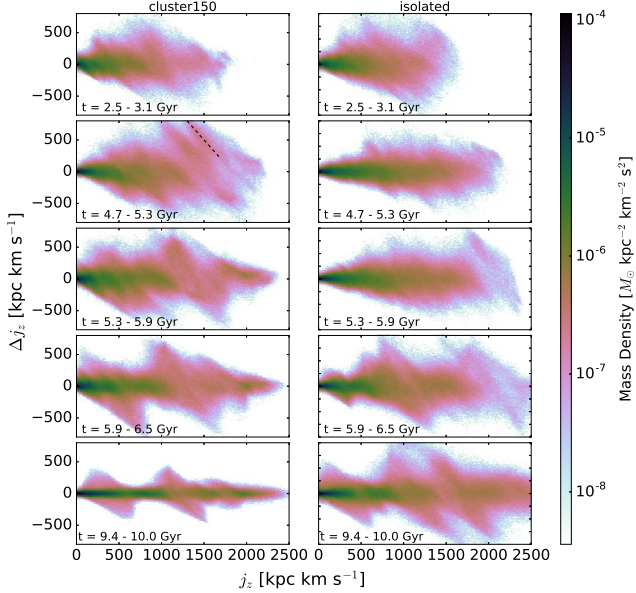}
  \caption{Mass-weighted distributions of $\Delta j_z$ given starting
    $j_z$ for the cluster150 model (left) and isolated model (right)
    at different times. The time intervals at which $\Delta j_z$ is
    computed is indicated at the lower left of each panel. The black dashed
    line (left column, second row) indicates the feature
    described in the text.}
\label{fig:jz}
\end{figure}

Without heating to move stars from the inner to the outer disc, we
consider whether the transient spiral migration mechanism of
\citet{sellwood2002} can explain the presence of these stars in the
outer disc.  In Figure~\ref{fig:jz} we show the change in angular
momentum for stars at different times in the cluster150 model (left
column) and in the isolated model (right column).  In each panel,
strong spiral-driven migration manifests as lines of negative gradient
\citep{sellwood2002, roskar2012}.  Particles with substantial positive
$\Delta j_z$ have moved outward, whilst particles with negative
$\Delta j_z$ have migrated inward.  The top row, at $t=2.5-3.1$~Gyr,
before the onset of significant RPS, shows that at
this time the cluster150 and isolated models exhibit grossly similar
spiral migration characteristics. By $t=4.7-5.3$~Gyr, during which
time the galaxy passes periapsis, the isolated150 model exhibits the
traces of strong, tidally-induced spirals in the outer disc.  Because
these spirals are in the outer disc, there is a substantial asymmetry
between outwards and inwards migrators, with the net result of a
strong outward migration into the outer disc that is not present in
the isolated model.  The outermost feature at this time (indicated by
the green dashed line) corresponds to
a circular angular momentum $j_c \sim 1850$~kpc~km~s$^{-1}$ or a
corotation radius of $\sim 8$~kpc; a $\Delta j_z \sim
600$~kpc~km~s$^{-1}$ will then move stars out to a radius of $ \sim
12$~kpc. At this time, this is the extreme outer disc, so migration by
tidally-induced spirals is likely responsible for populating the
outer disc.
At later times spirals in the cluster150 model slowly die out as the
quenching of SF robs the disc of the kinematically cool
stellar populations needed for sustaining spiral activity
\citep{sellwood1984}.  In contrast, SF continues in the
isolated model, allowing the disc to grow and spiral activity to
persist, permitting its outer disc mass to eventually catch up with
that of the cluster150 model.

We also examined similar plots for the cluster350 model, and found
that strong, tidally-induced outward migrating features are not
present to the same extent.  Instead this model is not fully quenched
and the continuing SF at intermediate radii feeds the
outer disc via the usual spiral migration mechanism.

\section{Observational consequences}

\begin{figure}
  \centering
  \includegraphics[width=0.99\columnwidth]{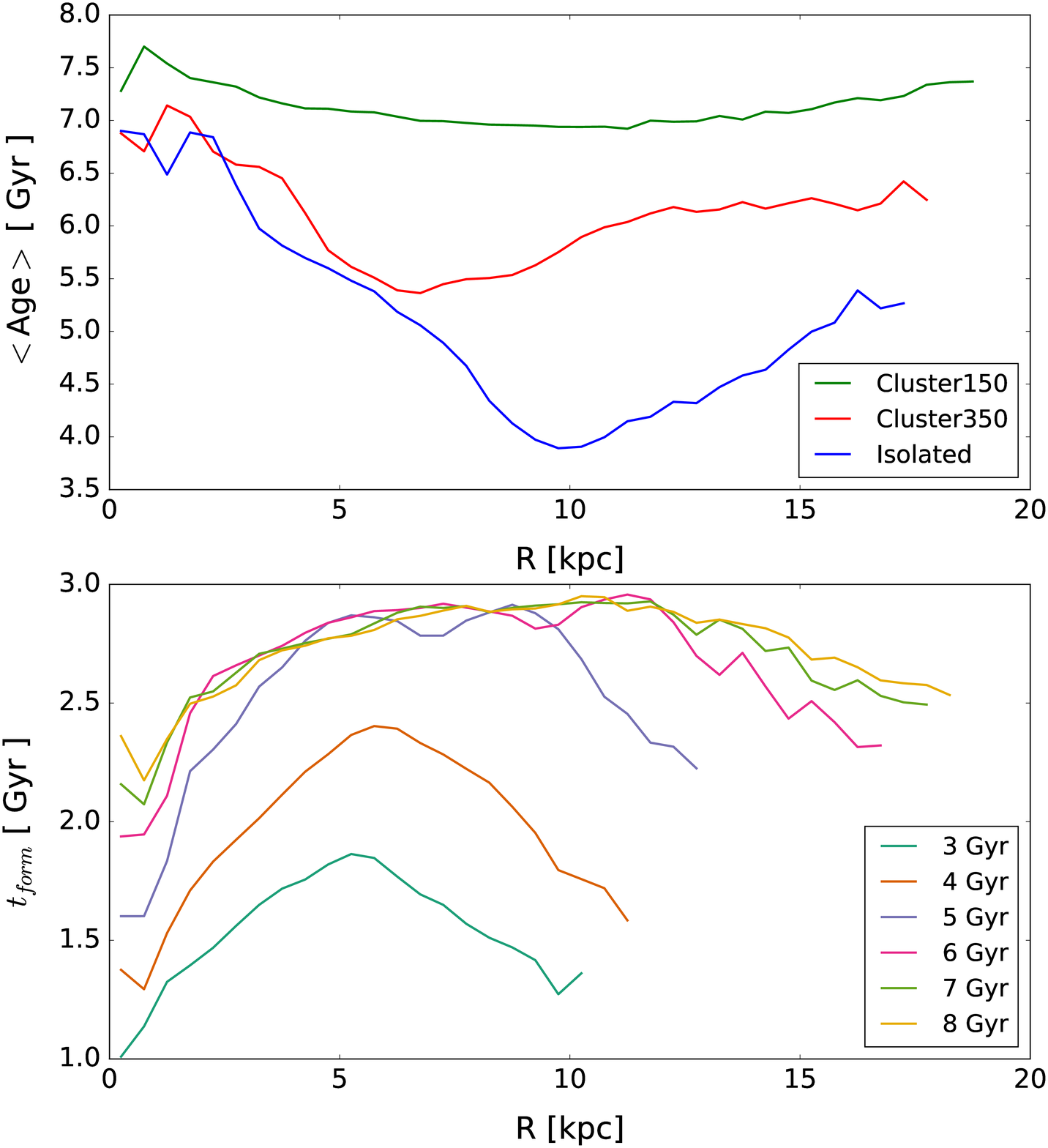}
  \caption{{\it Top\/}: Mean age profiles for our models at
  t=10~Gyr as detailed by the inset. {\it Bottom\/}: Profiles of the
  evoluton of the average formation time of stars for the cluster150
  model.}
\label{fig:age_feh}
\end{figure}

In the top panel of Figure~\ref{fig:age_feh} we plot age profiles for
the cluster models and the isolated model. As shown by
\citet{roskar2008a}, the age profile for the isolated model shows an
upturn in average age past the break due to the migration of older
stars into the outer disc.  In comparison, we find that the cluster150
model shows a flatter average age across the entire disc.  The bottom
panel shows the evolution of the formation time profile for the cluster150 model
in more detail. Until the RPS becomes efficient,
the profile has the distinctive down-turn expected from the
inside-out growth plus migration mechanism \citep{roskar2008b}. Once
the quenching and tidally-induced migration occurs, it
rapidly becomes flatter and remains this way thoughout.  Instead the
cluster350 model exhibits an age minimum at 6 kpc, which is the
maximum extent of the high SFR region (though at lower
levels SF proceeds beyond this point even to late times).

\section{Conclusions}

Using $N$-body+SPH simulations of galaxies falling into a gas-rich
cluster environment, we have shown that gas stripping and tidally-induced
spirals cause a transition from a type~II to a type~I
profile. Evolved in isolation, the model galaxy develops a type~II
profile, whilst the cluster galaxies exhibit either a weakening of the
break, or a profile close to that of a type~I, depending on the
orbital parameters of the galaxy. The radial velocity dispersion is not
increased by enough to
account for radial excursions of $\Delta R > 10$~kpc needed to
populate the outer disc. Instead, we show that there is an increase in
spiral activity in the outer disc induced by the cluster potential,
causing large outward radial migrations.  This efficiently
redistributes the material from the inner to the outer disc, while
retaining nearly circular
orbits~\citep[e.g.][]{sellwood2002,roskar2012}.
Whilst observationally few type~II profiles are detected
in cluster lenticulars~\citep{gutierrez2011,erwin2012,roediger2012},
our models retain breaks of varying strength. The cluster simulations
presented here have considered only one galaxy in the cluster at a
time, whilst in real clusters many galaxies will be present.  High-speed
tidal interactions between pairs of galaxies may also excite strong
spirals in the outer disc \citep[e.g.][]{moore1996, moore1999},
providing the possibility for type~I profiles to form in
galaxies that do not reach the cluster core, as long as RPS
is still able to quench.
This can explain why type~I profiles are more common amongst cluster
lenticulars, but does not explain how they occur in the field.  The
origin of type~I profiles amongst field spiral and lenticular galaxies
remains a puzzle.  We speculate that interactions between galaxies
play a role in the formation of type~I profiles in field lenticulars
too, together with quenching, which however may not be due to ram
pressure stripping.  Whether interactions play any role in type~I
profiles amongst spiral galaxies remains less clear.  NGC~300 is a
famous example of a nearby type~I spiral galaxy
\citep{bland-hawthorn2005}.  Integrated stellar populations from
broadband colours \citep{munoz-mateos+07} and {\it Hubble Space
  Telescope\/} resolved stellar populations \citep{gogarten+10} find
evidence of an inside-out growth, with relatively old ages.
\cite{bland-hawthorn2005} show the outer disc has
Toomre $Q = 5 \pm 2$, making it unfavourable to strong self-excited
spiral structure.  Together with its low mass, this makes migration in
NGC~300 inefficient \citep{gogarten+10}.~\cite{minchev2011} invoked
NGC~300 as an example of very efficient migration driven by resonance
overlap.  However the presence of radial metallicity gradients, in old
stars \citep{gogarten+10}, makes it unlikely that it has experienced
extreme migration. Instead type~I profiles may be the result
of a narrow range of halo angular momenta~\citep{herpich2015}.

Finally, we have shown that in lenticular galaxies the age profile
does not show the large upturn expected from the SF
threshold and migration mechanism \citep{roskar2008b}.  Instead the
age distribution is quite flat outside the region corresponding
roughly to the break radius at the time of quenching, while inside
this region a negative age gradient is present, which is the usual
signature of inside out growth.  This provides a testable prediction
for our scenario for the formation of type~I profiles in cluster
lenticulars.

\section*{Acknowledgements}
The simulations in this paper were run on the COSMOS Shared Memory system at
DAMTP, University of Cambridge operated on behalf of the STFC DiRAC HPC
Facility. This equipment is funded by BIS National E-infrastructure capital
grant ST/J005673/1 and STFC grants ST/H008586/1, ST/K00333X/1. VPD is
supported by STFC Consolidated grant \#~ST/M000877/1.
%%%%%%%%%%%%%%%%%%%%%%%%%%%%%%%%%%%%%%%%%%%%%%%%%%



\bibliographystyle{mnras}
\bibliography{library.bib}

%%%%%%%%%%%%%%%%%%%%%%%%%%%%%%%%%%%%%%%%%%%%%%%%%%

% Don't change these lines
\bsp% typesetting comment
\label{lastpage}
\end{document}